\documentclass[sn-mathphys,Numbered]{sn-jnl}

\usepackage{graphicx}%
\usepackage{multirow}%
\usepackage{amsmath,amssymb,amsfonts}%
\usepackage{amsthm}%
\usepackage{mathrsfs}%
\usepackage[title]{appendix}%
\usepackage{xcolor}%
\usepackage{textcomp}%
\usepackage{manyfoot}%
\usepackage{booktabs}%
\usepackage{algorithm}%
\usepackage{algorithmicx}%
\usepackage{algpseudocode}%
\usepackage{listings}%
\usepackage{float}
\usepackage{hyperref}

\theoremstyle{thmstyleone}%
\theoremstyle{thmstyletwo}%

\theoremstyle{thmstylethree}%

\begin{document}

\title[Article Title]{Compressive Fourier-Domain Intensity Coupling (C-FOCUS) enables near-millimeter deep imaging in the intact mouse brain in vivo}


\author[1]{\fnm{Renzhi} \sur{He}}\email{cubhe@ucdavis.edu}
\equalcont{These authors contributed equally to this work.}

\author[1]{\fnm{Yucheng} \sur{Li}}\email{ycsli@ucdavis.edu}
\equalcont{These authors contributed equally to this work.}

\author[2]{\fnm{Brianna} \sur{Urbina}}\email{bmurbina@ucdavis.edu}

\author[2]{\fnm{Jiandi} \sur{Wan}}\email{jdwan@ucdavis.edu}

\author*[1]{\fnm{Yi} \sur{Xue}}\email{yxxue@ucdavis.edu}

\affil[1]{\orgdiv{Department of Biomedical Engineering}, \orgname{University of California, Davis}, \city{Davis}, \postcode{95616}, \state{CA}, \country{United States}}

\affil[2]{\orgdiv{Department of Chemical Engineering}, \orgname{University of California, Davis}, \city{Davis}, \postcode{95616}, \state{CA}, \country{United States}}


\abstract{Two-photon microscopy is a powerful tool for in vivo imaging, but its imaging depth is typically limited to a few hundred microns due to tissue scattering, even with existing scattering correction techniques. Moreover, most active scattering correction methods are restricted to small regions by the optical memory effect. Here, we introduce compressive Fourier-domain intensity coupling for scattering correction (C-FOCUS), an active scattering correction approach that integrates Fourier-domain intensity modulation with compressive sensing for two-photon microscopy. Using C-FOCUS, we demonstrate high-resolution imaging of YFP-labeled neurons and FITC-labeled blood vessels at depths exceeding 900 $\mu$m in the intact mouse brain in vivo. Furthermore, we achieve transcranial imaging of YFP-labeled dendritic structures through the intact adult mouse skull. C-FOCUS enables high-contrast fluorescence imaging at depths previously inaccessible using two-photon microscopy with 1035 nm excitation, enhancing fluorescence intensity by over 20-fold compared to uncorrected imaging. C-FOCUS provides a broadly applicable strategy for rapid, deep-tissue optical imaging in vivo.}





\maketitle

\section{Introduction}\label{sec1}

Two-photon microscopy is one of the most widely adopted techniques for in vivo deep brain imaging at subcellular resolution. State-of-the-art two-photon systems incorporating adaptive optics for aberration correction can achieve imaging depths of 500-800 $\mu$m using yellow-green fluorophores in the intact mouse brain through a cranial window or thinned skull \cite{Helmchen2005-lg,Ji2017-vf, Rodriguez2021-sy, Chen2021-kq, Drew2010-bp, Papadopoulos2016-sy, Takasaki2020-od, Mittmann2011-jo}. The primary goal of adaptive optics with phase modulation is to improve resolution by correcting aberrations and/or scattering, which in turn enhances fluorescence intensity indirectly. A critical component of phase-based adaptive optics is the correction phase mask, which can be generated via direct wavefront sensing \cite{Cha2010-qe, Aviles-Espinosa2011-vt, Wang2014-wo,Wang2015-fl, Liu2019-ae, Chen2021-kq} or by indirectly measuring the aberrated point spread function (PSF) using interference or virtual guide stars \cite{Albert2000-oj,Marsh2003-rh, Debarre2009-ax, Rueckel2006-lb, Tang2012-bl, Papadopoulos2016-sy, May2021-er, Rodriguez2021-sy, Qin2022-ns}. Both direct and indirect approaches rely on ballistic photons, and thus the maximum imaging depth after correction remains fundamentally limited by photon scattering. Additionally, correction speed is constrained by the projection rate of spatial light modulators (SLMs) and the iterative process of phase retrieval. Due to the optical memory effect \cite{Freund1988-va}, which restricts the effective range of a single correction mask (``isoplanatic patch size"), there is a trade-off between field of view (FOV) and total correction time, which scales with the number of required isoplanatic patches \cite{Park2017-ol,May2021-hh, Blochet2023-ot}. 

Instead of phase modulators, intensity modulators have recently been employed for aberration \cite{Ren2020-xy, Chen2020-qv} and scattering correction \cite{Zepeda2025-cm} in two-photon microscopy, enabled by the availability of high-power femtosecond lasers. A digital micromirror device (DMD) is a commonly used intensity modulator that offers a patterning rate of 4 to 32.5 kHz, making it over ten times faster and more cost-effective than state-of-the-art SLMs. Recently, we developed 2P-FOCUS \cite{Zepeda2025-cm} that uses a DMD to correct scattering for deep tissue imaging. Unlike adaptive optics with phase modulation, 2P-FOCUS directly enhances fluorescence intensity by leveraging constructive interference among multi-scattered beams rather than relying on ballistic photons. 2P-FOCUS selectively reinforces in-phase beam combinations that contribute to constructive interference, guided by the resulting two-photon excited fluorescence from all possible multiple-beam interactions. We have demonstrated the effectiveness of 2P-FOCUS by ex vivo imaging of brain tissue.

Despite recent progress, two-photon microscopy systems with intensity-based scattering correction using a DMD have not yet been demonstrated for in vivo brain imaging. This is primarily due to the need for further improvements in correction speed to enable multipatch, multi-plane correction over sufficiently large volumes while avoiding motion artifacts and dynamic scattering during live imaging. As imaging depth increases, the optical memory effect range in the lateral dimension decreases significantly, often shrinking to just tens of microns \cite{Yoon2020-sm, Zepeda2025-cm}, which makes multipatch correction essential for a FOV larger than 100 $\mu$m. In the axial direction, a single correction mask is ineffective across multiple $z$-planes separated by the depth-of-field, as it cannot correct spatially varying scattering at different depths but instead only provides a defocused image of the corrected plane. As a result, the total correction time scales rapidly with imaging volume and can exceed one hour for volumetric datasets. More importantly, to achieve imaging depths beyond the 500-800 $\mu$m depth limit of the state-of-the-art two-photon microscopy with yellow-green fluorophores \cite{Takasaki2020-od, Mittmann2011-jo, Rodriguez2021-sy, Papadopoulos2016-sy, Chen2021-kq}, the effectiveness of scattering correction must also be improved. We found higher correction fidelity can be achieved by using phase masks with smaller super-pixels in 2P-FOCUS \cite{Zepeda2025-cm}, which provide more degrees of freedom but also require more raw measurements for proper sampling, further increasing the correction time. Therefore, to overcome the depth limit of current two-photon microscopy and enable deep in vivo brain imaging across large volumes, it is essential to improve both the speed and effectiveness of scattering correction.

To overcome these challenges of deep brain imaging in vivo, we propose \textit{c}ompressive \textit{Fo}urier-domain intensity \textit{c}o\textit{u}pling for \textit{s}cattering correction (C-FOCUS), a novel two-photon microscopy technique. C-FOCUS corrects scattering through intensity modulation on the Fourier plane using a DMD. It differs fundamentally from adaptive optics with phase modulation in three key ways:
(1) C-FOCUS is designed to enhance laser power at the focus through scattering tissue, rather than to improve resolution, while keeping the input power at the sample surface the same with or without correction; (2) it relies on multi-scattered photons instead of ballistic photons, thereby overcoming the limitations of exponentially decayed ballistic photons; and (3) it enhances constructive multi-beam interference rather than relying on phase conjugation. C-FOCUS improves upon our previous method, 2P-FOCUS \cite{Zepeda2025-cm}, by incorporating compressive sensing, which reduces the required measurements to 20-30\% and results in a 5-fold increase in overall acquisition and reconstruction speed. This enables C-FOCUS for in vivo imaging since it is able to complete the full correction workflow, including data acquisition, transfer, computation, and projection, within 1.5 minutes for tens of subregions across a $126 \times 126~\mu m^2$ FOV. Furthermore, its correction effectiveness is enhanced by generating masks with more super-pixels, providing greater degrees of control over the illumination pattern to correct highly spatially varying scattering. Using C-FOCUS, we achieve in vivo imaging depths beyond 900 $\mu$m in the intact mouse brain with subcellular resolution, resolving fine structures such as axons in layer 6 of the cortex. We also demonstrate transcranial imaging of neuronal dendrites through the intact skull---both capabilities not previously achieved with two-photon excitation at 1035 nm.

\section{Results}\label{sec2}
\subsection{The principle of C-FOCUS}\label{subsec2.1}

\begin{figure}[ht]%
\centering
\includegraphics[width=1\textwidth]{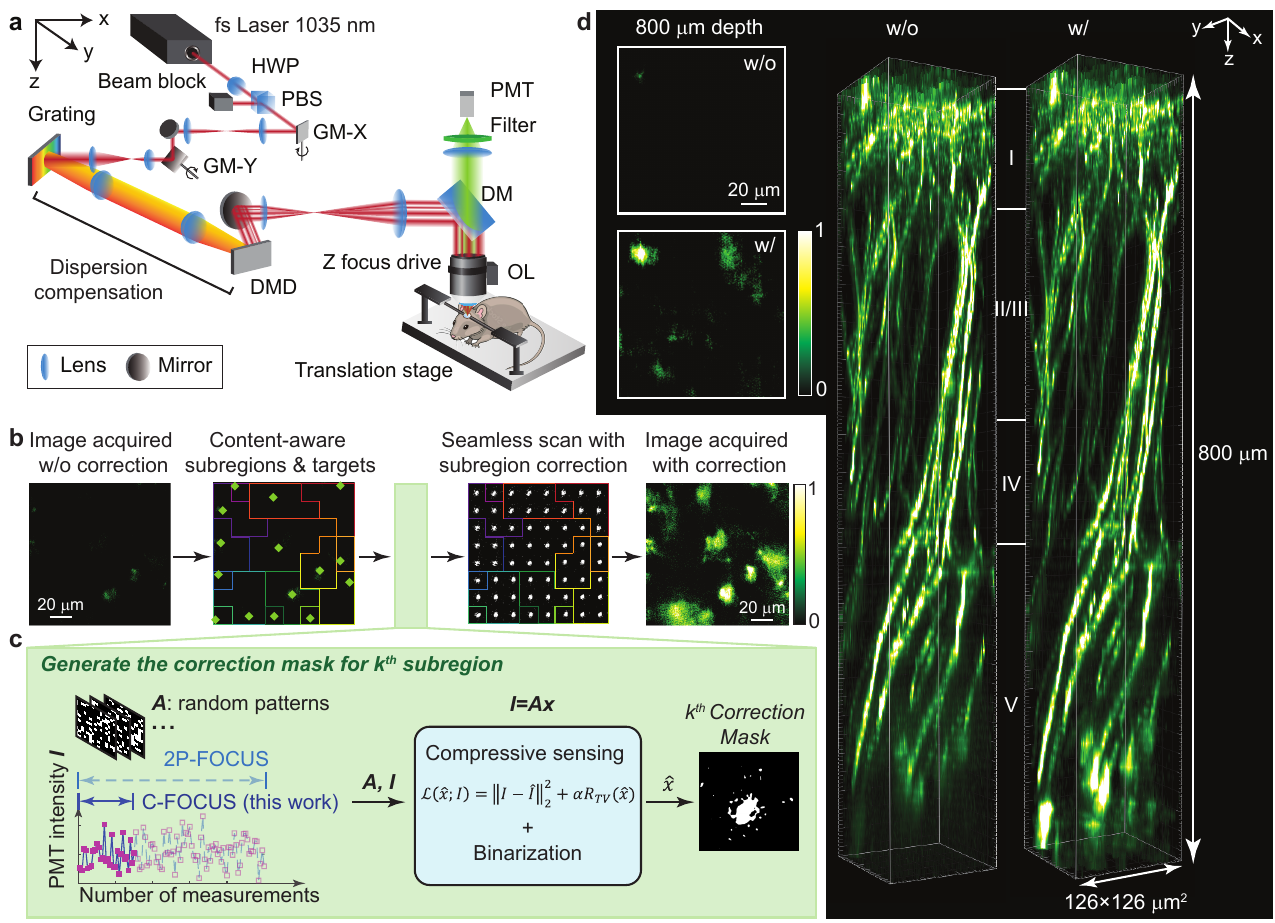}
\caption{\textbf{C-FOCUS enables two-photon imaging of YFP-labeled pyramidal neurons up to 800 $\mu m$ below the dura in the somatosensory cortex using intensity-based scattering correction with compressive sensing. a.} Optical schematic of the C-FOCUS system. Details are provided in \hyperref[Methods]{Methods}. \textbf{b-c.} Correction process of C-FOCUS: \textbf{b.} an uncorrected image is first acquired and segmented into subregions, each containing a local intensity peak that serves as a correction target. Correction masks are generated for these content-aware subregions and synchronously projected as the galvo mirrors scan, enabling image acquisition with active scattering correction. \textbf{c.} Correction masks for the subregions are measured and calculated sequentially using compressive sensing. \textbf{d.} In vivo imaging of YFP-labeled pyramidal neurons in the somatosensory cortex of a Thy1-YFP-H mouse through a cranial window using 1035 nm excitation, with and without scattering correction. With C-FOCUS, fluorescence intensity is enhanced 5-fold, enabling imaging to a depth of 800 $\mu$m (8.5 EAL).}\label{fig1}
\end{figure}

The optical setup of C-FOCUS is adapted from 2P-FOCUS \cite{Zepeda2025-cm} and optimized for improved performance (Fig.~\ref{fig1}a). It is a point-scanning two-photon microscopy system with a DMD placed at the relayed pupil plane for scattering correction. A diffraction grating is used to compensate for the spatial dispersion introduced by the DMD. In all experiments, the excitation wavelength is 1035 nm with a 1 MHz repetition rate. Point-scanning images are acquired at 288 $\times$ 288 pixels using galvanometric mirrors at 4 kHz scanning rate. The numerical aperture (NA) of the objective lens is 1.0. Further details of the optical setup, image acquisition, and image processing are in the \hyperref[Methods]{Methods}. To correct scattering at a specific depth (Fig.~\ref{fig1}b), C-FOCUS first acquires an image of the full FOV without correction by projecting a blank mask. The image is then segmented into content-aware subregions (see \hyperref[Methods]{Methods}) that adapt to object boundaries and extend the FOV beyond the memory effect range, preserving structural features more effectively than uniform grid-based patches \cite{Park2017-ol, Blochet2023-ot}. As an example (Fig.~\ref{fig1}b), an image of YFP-labeled neurons in the visual cortex acquired at 800 $\mu$m depth in vivo using 12 mW power on the brain surface is segmented into 15 content-aware subregions. The use of content-aware subregions significantly accelerates the multi-subregion correction process by avoiding redundant generation of identical correction masks and eliminating ineffective masks that may results from grid-based segmentation.

Next, C-FOCUS generates a correction mask for each subregion (Fig.~\ref{fig1}c). The core principle of Fourier-domain intensity-based scattering correction is to identify in-phase beams that constructively interfere at the target, enhancing the main lobe of the scattered speckle, while blocking out-of-phase beams that destruct interference. These in-phase beams do not need to be ballistic photons; they may be multiply scattered as long as they follow similar optical paths and maintain phase coherence. To probe the optimal multiple-beam interference at the target, C-FOCUS excites the target by projecting orthogonal random binary patterns ($A$) in the Fourier domain while recording the corresponding fluorescence intensity ($I$) using a photomultiplier tube (PMT). These random patterns generate multiple narrow beams that may or may not constructively interference after propagating through scattering tissue, as reflected by fluctuations in the measured fluorescence intensity (Fig.~\ref{fig1}c, left). Constructive interference yields high fluorescent intensity, whereas destructive interference generates lower intensity. While 2P-FOCUS requires well-sampled measurements to resolve the contribution of each superpixel to constructive interference, C-FOCUS only requires 20\%-30\% of those measurements by leveraging compressive sensing. The correction mask ($\hat{x}$) is estimated by solving the inverse problem with a total variation (TV) regularizer:
\begin{equation}
\mathcal{L}(\hat{x}; I) =  \|I-A\hat{x}\|_2^2 + \alpha  R_{TV}(\hat{x}),
\label{eq1}
\end{equation}
followed by binarization. Details regarding the number of required measurements and the influence of hyperparameters are discussed in Section \ref{subsec2.2}, and details on computing correction masks are provided in \hyperref[Methods]{Methods}. C-FOCUS is also a single-shot correction method that does not require iterative, hardware-involved optimization, which significantly improves speed compared to adaptive optics. A single set of random patterns is preloaded and reused across all subregions, minimizing data transfer time. For the representative example in Fig.~\ref{fig1}b, the complete generation of all 15 correction masks (each containing 10,000 superpixels) for a $126 \times 126~\mu m^2$ FOV takes less than 1.5 min (89.3 seconds), representing a 5-fold speed improvement over the estimated time required by 2P-FOCUS.

After generating all subregion correction masks, C-FOCUS ensures that the laser power delivered to the sample surface remains constant with and without correction by increasing the input power to the DMD (Details about the transmission efficiency is discussed in Section \ref{subsec2.2}). In the final step, the correction masks are projected while seamlessly scanning the entire FOV, achieved by synchronizing DMD projection with galvanometric mirror scanning, rather than scanning each subregion sequentially and stitching the images post hoc \cite{Wang2014-wo}. Compared to the image acquired without correction, the corrected image shows a substantial improvement in fluorescence intensity across the entire FOV (Fig.~\ref{fig1}b, right). 

C-FOCUS represents a substantial breakthrough in two-photon microscopy, enabling high-resolution imaging at depths that have traditionally posed significant challenges \cite{Helmchen2005-lg,Ji2017-vf, Takasaki2020-od, Wang2020-zb}. These limitations arise not only from the exponential attenuation of ballistic photons but also from the increased scattering in deeper brain regions, including deep cortical layers and white matter, and areas with dense vasculature. We quantitatively characterized the depth-dependent effective attenuation length (EAL) in vivo across different brain regions (Extended Data Fig.~\hyperref[secA1]{1}a-f). To demonstrate C-FOCUS, we imaged YFP-labeled pyramidal neurons in the somatosensory cortex of a Thy1-YFP-H mouse in vivo through a cranial window, using 68 mW excitation power at the brain surface. This region has dense vasculature, which causes severe scattering. With C-FOCUS, the imaging depth is extended to 800 $\mu$m (8.5 EAL), resulting in a 5-fold increase in fluorescence intensity and enabling clear visualization of neurons beyond the conventional depth limit (Fig.~\ref{fig1}d).

\subsection{Quantitative evaluation of C-FOCUS by imaging fluorescent beads through ex vivo mouse skull}\label{subsec2.2}

To quantitatively evaluate and optimize correction masks generated under various parameters, we imaged red fluorescent beads embedded in PDMS through an ex vivo mouse skull using these masks. The skull was approximately 250 $\mu$m thick, equivalent to 5.3 EALs (EAL = 47.2 $\mu$m, Extended Data Fig.~\hyperref[secA1]{1}c). Details on sample preparation are provided in \hyperref[Methods]{Methods}.

\subsubsection{Determining the minimal number of measurements for effective correction}

We investigated the minimal number of measurements required for C-FOCUS to generate effective correction masks. As C-FOCUS solves an ill-posed inverse problem from undersampled intensity measurements, increasing the number of measurements improves correction quality but also increases acquisition and computation time. As a baseline, we imaged a fluorescent bead without correction by projecting a blank screen on the DMD, resulting in low SNR, with the bead barely distinguishable from background speckle (Fig.~\ref{fig2}b). The laser power on the sample is 6 mW. Based on our previous 2P-FOCUS work \cite{Zepeda2025-cm}, correction is optimized using random patterns composed of 100$\times$100 superpixels (with $8 \times 8$ pixels per superpixel) at a sparsity of 0.4 (i.e., 40\% of superpixels turned on). Fully sampling the 10,000 unknown superpixels requires 10,000 measurements, whereas compressive sensing allows effective correction with fewer. We acquired 10,000 measurements and generated correction masks from subsets of 10 to 5,000 measurements, which were then applied to image the same bead and evaluate fluorescence intensity (Fig.~\ref{fig2}a). We also generated a correction mask using all 10,000 measurements with 2P-FOCUS and measured the resulting intensity (Fig.~\ref{fig2}d). The results show that fluorescence intensity increases with the number of measurements (Fig.~\ref{fig2}f), reaching a maximum 38.3-fold enhancement with 10,000 measurements. With C-FOCUS, the minimum improvement is 6.2-fold with only 10 measurements, and the maximum is 37.5-fold with 5,000 measurements. The intensity begins to plateau beyond 2,000 measurements, where over 85\% of the maximum enhancement has already been achieved. In addition, we compared the fluorescence intensity after correction using two different sets of 2,000 random patterns and found that the intensity difference was less than 5\% (Extended Data Fig.~\hyperref[secA2]{2}a-b), indicating that the results are largely independent of the choice of random pattern seeds. Furthermore, hyperparameters in the compressive sensing algorithm, such as the weight of the TV regularizer, may also slightly affect correction effectiveness (Extended Data Fig.~\hyperref[secA2]{2}e-f). In summary, C-FOCUS requires only about 20\% of the fully sampled measurements to achieve effective correction, making it well-suited for in vivo imaging applications.

To demonstrate that C-FOCUS corrects spatially varying scattering rather than merely reducing the effective NA, we imaged the fluorescent bead using a low-NA mask with a comparable size to the DC component of the correction masks (Fig.~\ref{fig2}c). Since on-axis (low-NA) photons experience less scattering than off-axis (high-NA) photons, reducing the NA can enhance signal intensity in deep tissue imaging \cite{Helmchen2005-lg, Tung2004-ew, Kondo2017-lh, Chen2021-kq}. As shown here, while the low-NA mask produced a modest increase in fluorescence intensity compared to the uncorrected case, the resulting signal remained significantly weaker than that achieved with C-FOCUS (Fig.~\ref{fig2}f). This highlights the important role of the high-frequency components in the correction masks for effectively correcting complex, spatially varying scattering. Further comparison of the correction masks generated by 2P-FOCUS and C-FOCUS reveals strong agreement in both the DC and some high-frequency components (Fig.~\ref{fig2}e). Additionally, we also verified that photobleaching was minimal by re-imaging the same fluorescent bead without correction and observed only a 7.7\% decrease in fluorescence intensity after completing all imaging tasks (Extended Data Fig.~\hyperref[secA2]{2}c-d). These findings demonstrate that C-FOCUS effectively corrects highly spatially varying scattering using intensity masks with broad spatial frequency content, rather than relying on a simple low-NA mask.

\subsubsection{Determining the optimal DMD output-to-input power ratio for effective correction and high-resolution imaging}

\begin{figure}[H]%
\centering
\includegraphics[width=1\textwidth]{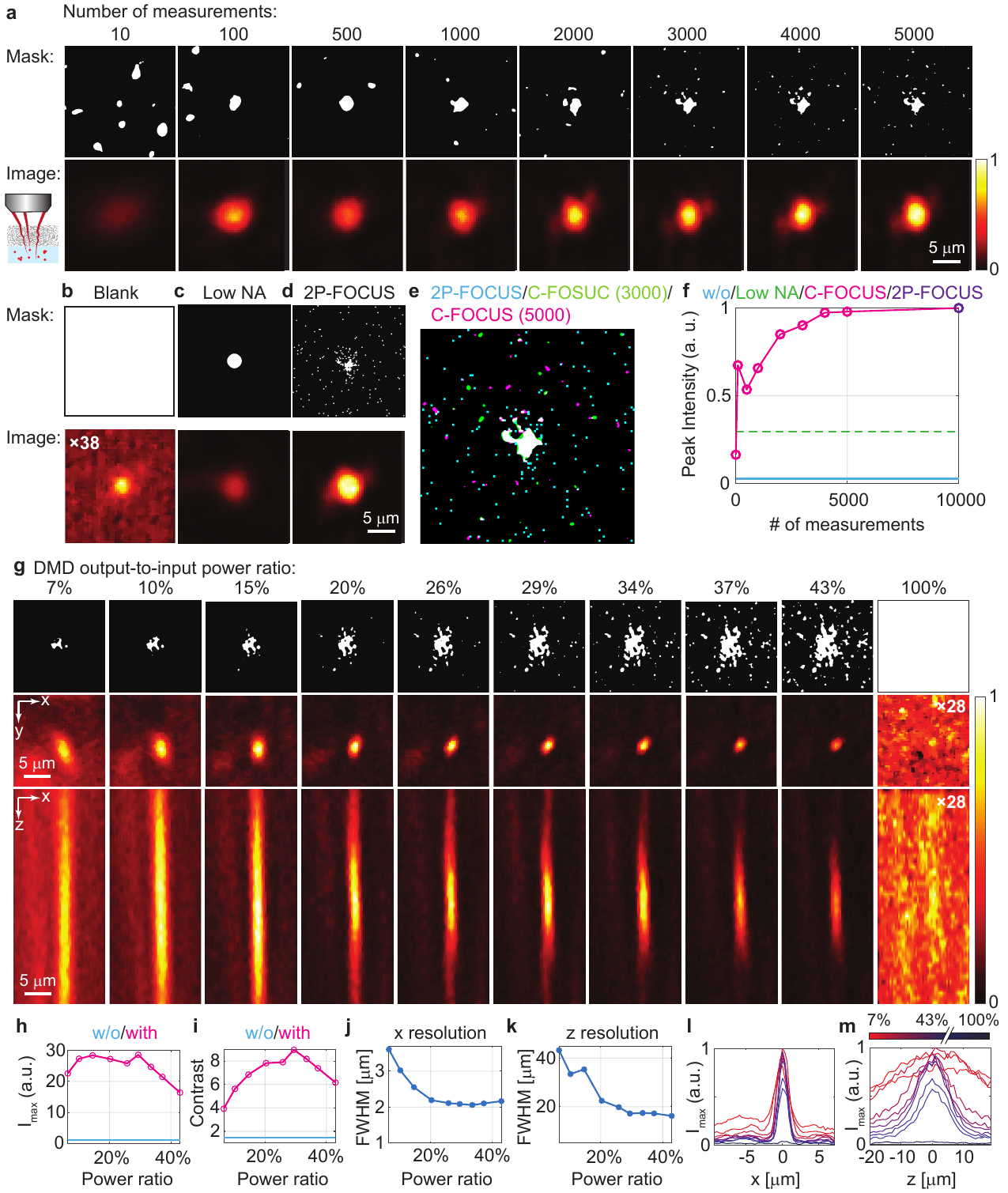}
\caption{\textbf{Quantitative evaluation of C-FOCUS by imaging red fluorescent beads through a 250 $\mu m$ mouse skull ex vivo. a.} Correction masks computed using 10-5,000 measurements and the corresponding corrected images. The fully sampled case uses 10,000 measurements. \textbf{b.} Image of the same fluorescent bead without correction (magnified 38-fold for display). \textbf{c.} Image of the bead using a low numerical aperture (NA) mask with size comparable to the DC component of the correction mask. \textbf{d.} Image using a correction mask generated by 2P-FOCUS with 10,000 measurements. \textbf{e.} Overlap of correction masks from 2P-FOCUS (cyan, 10,000 measurements) and C-FOCUS (green, 3,000 measurements; magenta, 5,000 measurements). \textbf{f.} Comparison of peak fluorescence intensity without correction (cyan), with low NA (green), with 2P-FOCUS (purple), and with C-FOCUS (magenta) using correction masks from \textbf{a}. \textbf{g.} 3D point spread functions (PSFs) measured through a 250 $\mu$m-thick skull using different correction masks generated by applying various binarization thresholds (DMD output-to-input power ratio) to the same grayscale correction mask. \textbf{h.} Fluorescence intensity, \textbf{i.} image contrast, \textbf{j.} lateral ,and \textbf{k.} axial resolution at various power ratios. \textbf{l, m.} PSF cross-sections of the bead along the $x-$ and $z-$axes, respectively.}\label{fig2}
\end{figure}

Another important parameter for generating an effective correction mask is the binarization threshold, which controls the output-to-input power ratio on the DMD, the effective NA, and the selection of in-phase beams that contribute to constructive interference through scattering media. A higher threshold selects superpixels with greater weights, meaning the corresponding beams are more likely to be in phase and enhance scattering correction. In contrast, a lower threshold includes superpixels associated with partially in-phase or out-of-phase beams, which contribute less or even negatively to constructive interference, as discussed in 2P-FOCUS \cite{Zepeda2025-cm}. While a high threshold improves scattering correction and yields a brighter, higher-resolution focus, setting the threshold too high significantly reduces the effective NA. While slightly reducing the NA can aid deep tissue imaging \cite{Helmchen2005-lg, Tung2004-ew, Kondo2017-lh}, an excessively low NA reduces both resolution and peak intensity. To determine the optimal point, we imaged a single fluorescent bead (0.71 $\mu$m) through the same mouse skull (Fig.~\ref{fig2}g) using correction masks generated with varying binarization thresholds. Without correction (output-to-input power ratio = 100\%), the fluorescent bead was indistinguishable from background speckle. As the output-to-input power ratio decreases, fewer superpixels were turned on in the correction masks (top row, Fig.~\ref{fig2}g). The combined effects of scattering correction and NA modulation resulted in increased intensity and resolution as the ratio decreased from 43\% to 29\%, followed by a decline from 29\% to 7\% (Fig.~\ref{fig2}g-m). Thus, the optimal DMD output-to-input power ratio is 29\%, which improves fluorescence intensity by 28.6-fold while maintaining a resolution of $2.1 \times 2.1 \times 17.2~\mu m^3$ when imaging through a 250 $\mu$m-thick mouse skull.

\subsection{C-FOCUS enables high-resolution in vivo imaging of pyramidal neuron axons up to 900 $\mu$m deep in layer 6 and white matter}\label{subsec2.3}

\begin{figure}[h!]%
\centering
\includegraphics[width=1\textwidth]{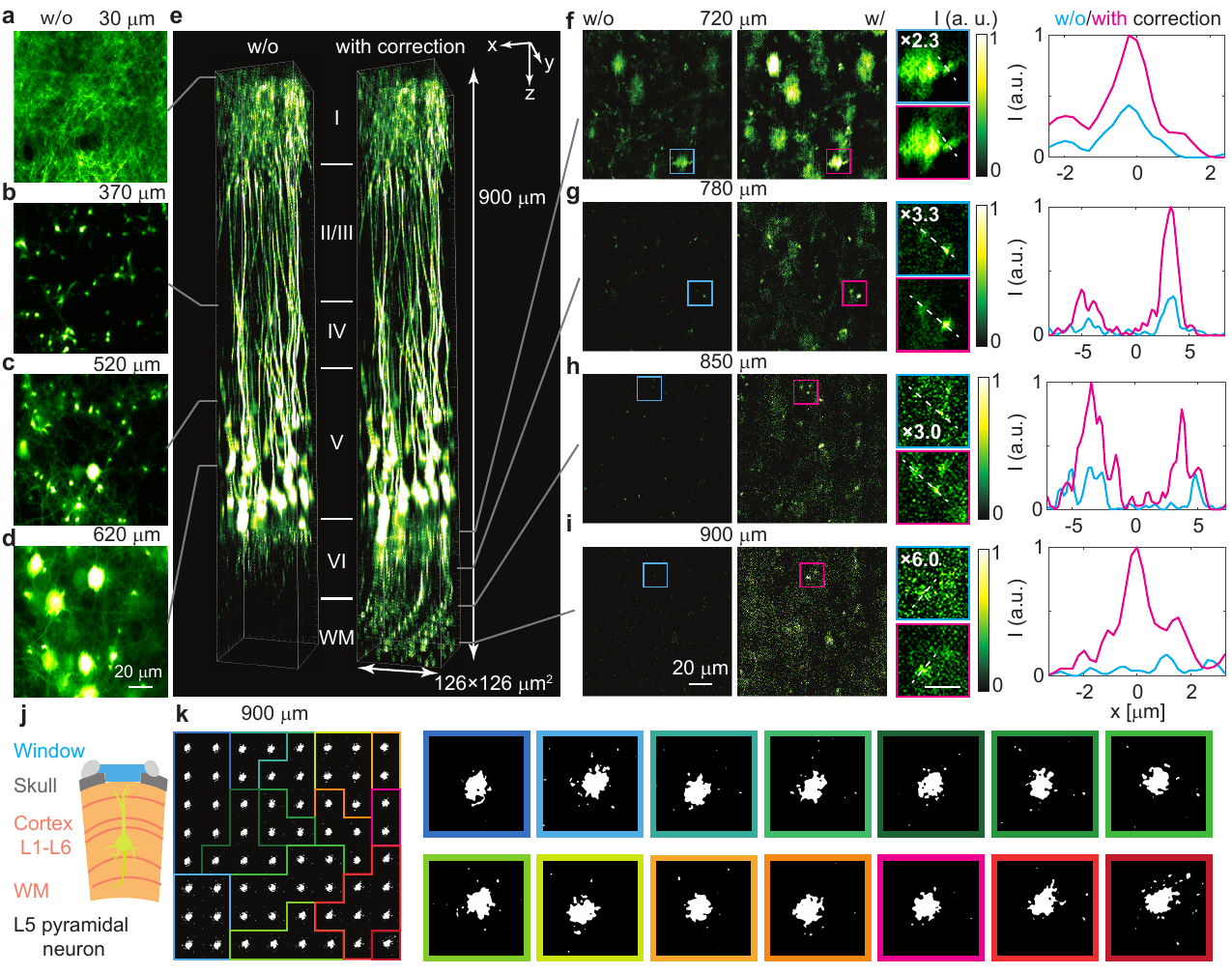}
\caption{\textbf{C-FOCUS visualizes axons of pyramidal neurons up to 900 $\mu$m deep in layer 6 and white matter of the visual cortex in vivo.} An adult transgenic Thy1-YFP-H mouse was imaged through a cranial window under anesthesia. \textbf{a-d.} Representative images at depths from 30 to 620 $\mu$m without correction. \textbf{e.} Comparison of volumetric views of images without correction (left) and with correction (right). \textbf{f-i} Side-by-side comparison of images without correction (first column), with correction (second column), zoomed-in views (third column; scale bar, 10 $\mu$m), and fluorescence intensity profiles along the dashed lines shown in the zoomed-in views (fourth column), at \textbf{f.} 720 $\mu$m, \textbf{g.} 780 $\mu$m, \textbf{h.} 850 $\mu$m, and \textbf{i.} 900 $\mu$m depths. In the zoomed-in views, intensities of the uncorrected images are magnified as indicated for visualization purposes. \textbf{j} Schematic of in vivo imaging of layer 5 pyramidal neurons expressing YFP through a cranial window. \textbf{k.} Subregions and corresponding correction masks at 900 $\mu$m depth, labeled by colored frames.}\label{fig3}
\end{figure}

We next apply C-FOCUS for high-resolution, deep brain in vivo imaging. We imaged YFP-labeled pyramidal neurons in the visual cortex of transgenic Thy-YFP-H mice through a cranial window (Fig.~\ref{fig3}j). Details on animal preparation are provided in \hyperref[Methods]{Methods}. Without correction, we reached a depth of approximately 720 $\mu m$ (Fig.~\ref{fig3}a-d, e, left), clearly visualizing tuft dendrites (representative image at 30 $\mu$m, Fig.~\ref{fig3}a), apical dendrites (representative image at 370 $\mu$m, Fig.~\ref{fig3}b), basal dendrites and soma (representative images at 520 $\mu$m and 620 $\mu$m, Fig.~\ref{fig3}c-d). The maximum laser power at the sample was 15 mW. At $z$-planes deeper than 720 $\mu$m, the brain tissue becomes significantly more scattering, as it includes the bottom of the cortex and the white matter, where the EAL decreases from 207.5 $\mu$m to 61.4 $\mu$m (Extended Data Fig.~\hyperref[secA1]{a-b}). As a result, fluorescence intensity drops rapidly with increasing depth, and no structures are visible without correction (Fig.~\ref{fig3}e, left; Fig.~\ref{fig3}f-i, left). 

Scattering correction was then performed starting at 720 $\mu$m and continuing at every 10 $\mu$m $z$-plane down to 900 $\mu$m. For each subregion, a correction mask was generated using 3,000 random patterns (40\% sparsity, 8 $\times$ 8 pixel superpixels, $100 \times 100$ superpixels per mask), thresholded at a DMD output-to-input power ratio of 29\%. With correction, we successfully visualized axons of layer 5 pyramidal neurons extending vertically through layer 6, turning nearly horizontally as they enter the white matter near the hippocampus. (Fig.~\ref{fig3}e, right). Representative images from 720 to 900 $\mu$m demonstrate a clear improvement in fluorescence intensity, with up to a 6-fold increase at 900 $\mu$m depth (Fig.~\ref{fig3}f-i). Zoomed-in views and intensity profiles reveal that C-FOCUS resolves axons with diameters of 1.1-2 $\mu$m in layer 6 and white matter (Fig.~\ref{fig3}f-i). Each of the 14 correction masks generated at 900 $\mu$m depth contains unique spatial features (Fig.~\ref{fig3}k), enabling correction of highly spatially varying scattering across a $126 \times 126 ~\mu m^2$ FOV. The measurement time is 3 seconds per correction mask, and the computation time is 2.9 seconds per mask, resulting in a total correction time of 83.2 seconds for all 14 subregions. Additional details on the imaging parameters are provided in \hyperref[Methods]{Methods} and Supplementary information. In summary, C-FOCUS is the first two-photon microscopy technique to achieve in vivo imaging up to 900 $\mu$m (7.8 EAL) in the intact brain with single-axon resolution, enabling clear visualization of fine axonal structures in layer 6 and the highly scattering white matter. 

\subsection{In vivo imaging of vasculature up to 910 $\mu$m deep with C-FOCUS over a large FOV}\label{subsec2.4}

\begin{figure}[h!]%
\centering
\includegraphics[width=1\textwidth]{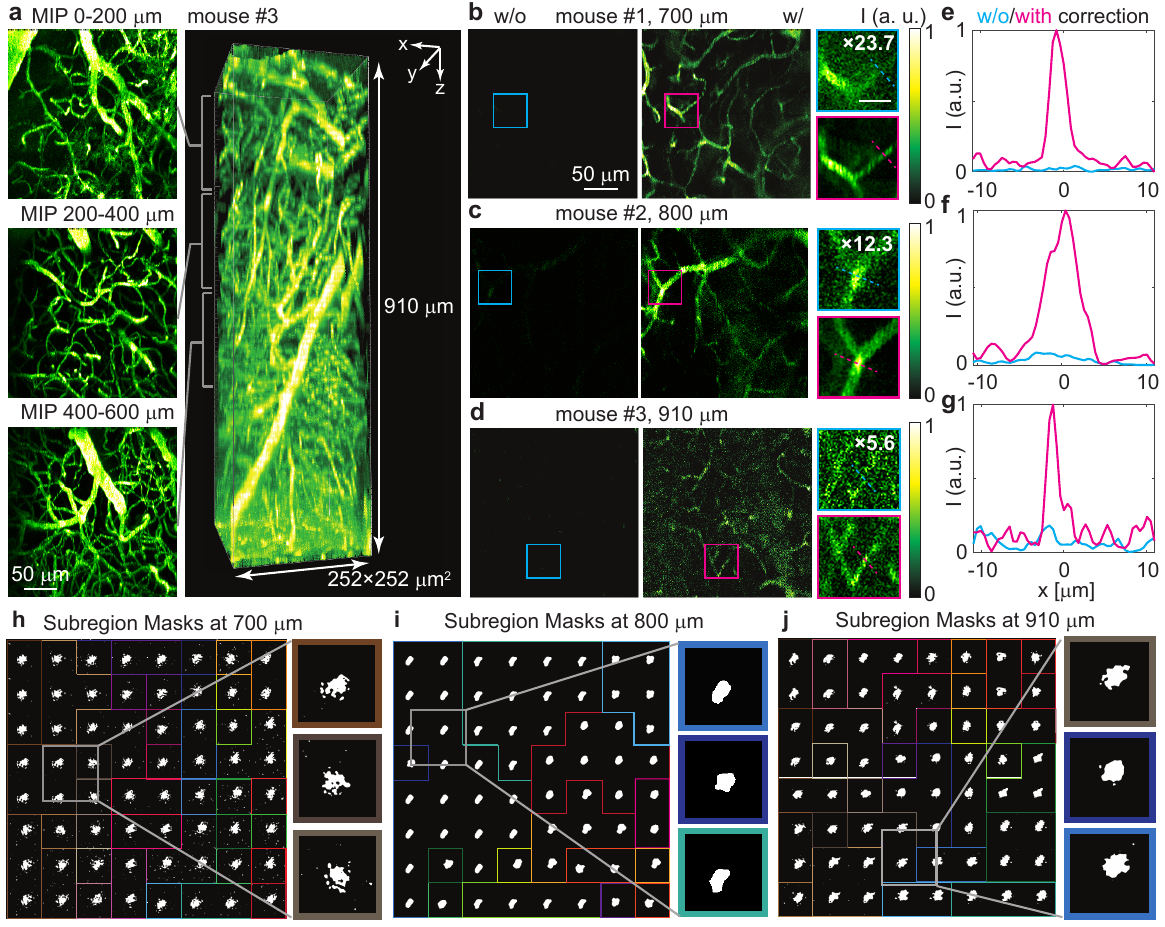}
\caption{\textbf{C-FOCUS visualizes FITC-labeled blood vessels up to 910 $\mu$m in vivo.} Adult wild-type mice were imaged through a cranial window under anesthesia following intravenous injection of FITC. \textbf{a.} Volume rendering of FITC-labeled blood vessels after scattering correction. The left column shows maximum intensity projection (MIP) images from 0 to 600 $\mu$m depth without correction. The right column shows the volume view of the $252 \times 252 \times 910 \mu m^3$ image stack. \textbf{b-d.} Comparison between images without correction (left column), with correction (middle column), and zoomed-in views (right column, scale bar: 20 $\mu$m) at \textbf{b} 700 $\mu$m, \textbf{c} 820 $\mu$m, and \textbf{d} 910 $\mu$m depths. These images were acquired from different mice under the same labeling protocol. In the zoomed-in views, the intensity of the uncorrected images is digitally magnified (scaling factor indicated in each panel) for display purposes. \textbf{e-g.} Fluorescence intensity profiles without and with correction along the dashed lines in the zoomed-in views. \textbf{h} Subregions and representative correction masks from the boxed regions in \textbf{b-d} (highlighted by colored frames).}\label{fig4}
\end{figure}

To evaluate C-FOCUS across a larger FOV with varying fluorescence types and labeling densities, we imaged blood vessels over a $252 \times 252 \times 910 \mu m^3$ volume in wild-type mice, labeled via intravenous injection of FITC and imaged through a cranial window in vivo. Details on animal preparation are provided in \hyperref[Methods]{Methods}. Without correction, we imaged blood vessels in the visual cortex with a maximum laser power of 74 mW on the sample surface. Maximum intensity projections (MIPs) over every 200 $\mu$m depth interval clearly reveal the pia arteries at the surface, penetrating arterioles, and a dense capillary network deep within the brain (Fig.~\ref{fig4}a). 

Next, we applied scattering correction using C-FOCUS with the same random patterns and binarization threshold as in the previous section. We imaged three different mice with the same labeling protocol and observed fluorescence intensity enhancements of 23.7-fold (mouse \#1 at 700 $\mu$m), 12.3-fold (mouse \#2 at 800 $\mu$m), and 5.6-fold (mouse \#3 at 910 $\mu$m) (Fig.~\ref{fig4}b-d). C-FOCUS also achieves high spatial resolution, resolving capillaries as small as 1.5 $\mu$m in diameter after correction at 910 $\mu$m depth (Fig.~\ref{fig4}e-g). The corresponding correction masks exhibit highly spatially varying features across subregions (Fig.~\ref{fig4}h-j). Compared to the axon imaging in the previous section, the scattering correction applied to blood vessel imaging yields significantly greater intensity improvements. This enhancement is largely due to the lower labeling density of FITC-labeled blood vessels compared to YFP-labeled pyramidal neurons. In the neuronal samples, both dendrites and axons express YFP, forming a dense network in layer 5 and deeper, which contributes to background fluorescence and reduces the contrast under random pattern modulation. The corrections at depths of 700 $\mu$m and 910 $\mu$m consist of more subregions (approximately 30) compared to neuronal imaging, which allows for more accurate correction across the larger FOV. The exposure time per random mask during measurement was also increased from 0.5 ms to 1-2 ms to improve the SNR of the raw measurements. Consequently, the total measurement and computation times increased to approximately 5 seconds and 17 seconds per correction mask, respectively. Additional details on the imaging parameters are provided in \hyperref[Methods]{Methods} and Supplementary information. In summary, these results demonstrate that C-FOCUS enables in vivo imaging of FITC-labeled blood vessels up to 910 $\mu$m (8.0 EAL) at 1035 nm excitation, over a 252$\times$252 $\mu m^2$ FOV and with 1.5 $\mu$m resolution. This performance surpasses the theoretical depth limit of conventional two-photon microscopy without correction and represents the first demonstration of such deep, high-resolution imaging at this wavelength.

\subsection{C-FOCUS enables in vivo visualization of apical dendrites through the intact mouse skull}\label{subsec2.5}

\begin{figure}[h!]%
\centering
\includegraphics[width=1\textwidth]{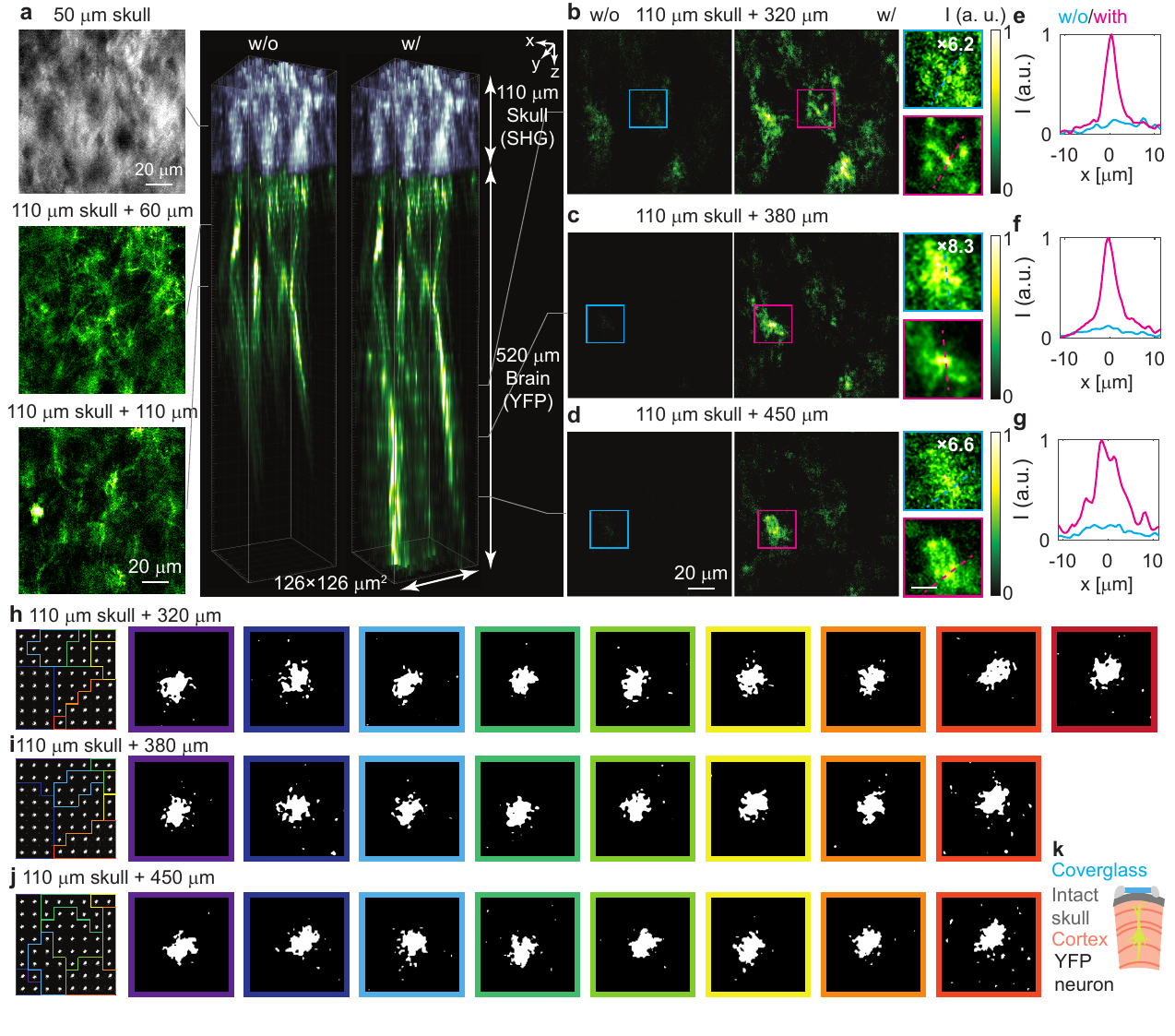}
\caption{\textbf{C-FOCUS visualizes the apical dendrites of YFP-labeled pyramidal neurons through 110 $\mu$m-thick intact skull in vivo.} Adult Thy-YFP-H mice is imaged through intact skull under anesthesia. \textbf{a.} The volume view of the skull (second-harmonic signal (SHG)) and YFP-labeled dendrites without and with correction. The left column shows the representative SHG image at 50 $\mu$m depth (1.1 EAL) and YFP images at 170 $\mu$m (3.2 EAL) and 220 $\mu$m (3.4 EAL) depth without correction. \textbf{b-d.} Comparison between the images without correct (left column) and the images with correction (middle column), and zoomed-in views (right column, scale bar, 10 $\mu$m) at \textbf{b} 430 $\mu$m (3.9 EAL), \textbf{c} 490 $\mu$m (4.2 EAL), and \textbf{d} 560 $\mu$m (4.5 EAL) depth. In the zoomed-in views, the intensity of images without correction are magnified (as labeled in the images) for display purpose. \textbf{e-g.} Fluorescent intensity along the dashed line in the zoomed-in views. \textbf{h-j} Subregions and correction masks on the same $z$-plane as \textbf{b-d}. \textbf{k.} Schematic of in vivo transcranial imaging of YFP neurons through an intact skull. The coverglass is to protect the exposed skull after survive surgery.}\label{fig5}
\end{figure}

We next demonstrate the capability of C-FOCUS in an even more challenging environment: in vivo imaging of YFP-labeled neurons through the intact mouse skull (Fig.~\ref{fig5}). While in vivo imaging through the intact skull has previously been achieved only with three-photon microscopy \cite{Wang2018-st, Qin2022-ns}, two-photon microscopy has not yet demonstrated this capability. Even with adaptive optics, two-photon imaging has only been shown through a thinned skull (about 50 $\mu$m on average) \cite{Drew2010-bp, Papadopoulos2016-sy, Chen2021-kq}, not an intact one. The skull is significantly more scattering than brain tissue, with an EAL of 46.5 $\mu$m in vivo (Extended Data Fig.~\hyperref[secA1]{1}e). We used a P54 (7.7-week-old) adult Thy1-YFP-H transgenic mouse for this experiment (Fig.~\ref{fig5}k). Without correction, we imaged the second-harmonic generation (SHG) signal from the 110 $\mu$m-thick (2.4 EAL) intact skull (gray-white region in Fig.~\ref{fig5}a), revealing structural features in the bone and marrow cavities. Below the skull, we captured images of tuft dendrites from YFP-labeled pyramidal neurons in the visual cortex without correction (Fig.~\ref{fig5}a), reaching a depth of 260 $\mu$m into the brain. 

We then applied scattering correction using C-FOCUS from 260 $\mu$m to 520 $\mu$m depth through the 110 $\mu$m-thick skull, corresponding to a total imaging depth of 4.9 EAL. The same random pattern set and binarization threshold from the previous section were used. The laser power on the sample was kept at 22 mW for both corrected and uncorrected images. With correction, we observed a 6.2- to 8.3-fold enhancement in fluorescence intensity while preserving high resolution, enabling visualization of apical dendrites with diameters of 3-3.5 $\mu$m (Fig.~\ref{fig5}b-c, e-f) as well as a soma at layer 5 (Fig.~\ref{fig5}d, g). The correction masks used for the 8-9 subregions at each imaging plane exhibited strong spatial variation and contained high-frequency features (Fig.~\ref{fig5}h-j). The measurement and computation times are 3 seconds and 2.5 seconds per correction mask, respectively, resulting in a total correction time of approximately 44 seconds per plane across 8-9 subregions. Additional details on the imaging parameters are provided in \hyperref[Methods]{Methods} and Supplementary information. Compared to apical dendrites imaged at similar depths through a cranial window (Fig.~\ref{fig3}a), the images through the intact skull displayed residual speckle patterns, even after correction. These could potentially be further improved with post-processing using a spatially varying estimated PSF. We also applied C-FOCUS to enhance SHG signals through a thick skull ex vivo, achieving correction down to 250 $\mu$m (5.4 EAL) (Extended Data Fig.~\hyperref[secA3]{3}). SHG signal density from the skull is much higher than that of both YFP-labeled pyramidal neurons and FITC-labeled blood vessels based on our measurements, which is more challenging for scattering correction. The result (Extended Data Fig.~\hyperref[secA3]{3}) indicates C-FOCUS can enhance SHG signals in highly scattering skull after scattering correction. 

Correcting scattering caused by the skull is significantly more challenging than correcting scattering in brain tissue, due to the skull’s highly heterogeneous structure and stronger spatial variation, which modulates light at higher spatial frequencies in three dimensions. For example, an image acquired through 2.4 EAL of cortical tissue (498 $\mu$m depth in the brain) can still exhibit high SNR and clearly resolved dendritic structures even without correction (Fig.~\ref{fig3}a). In contrast, imaging through 2.4 EAL of skull (110 $\mu$m thick) results in severe speckle artifacts and relatively low SNR, despite using higher laser power (3.6 mW vs. 7.3 mW). Despite the skull’s increased structural complexity, C-FOCUS enabled, for the first time, in vivo two-photon imaging of apical dendrites through the highly scattering intact mouse skull, achieving depths up to 630 $\mu$m (4.9 EAL) and overcoming a longstanding barrier in two-photon microscopy.

\section{Discussion}\label{sec12}

In this work, we present C-FOCUS, an active scattering correction method for two-photon microscopy that uses intensity modulation computed via compressive sensing. We demonstrate its capability through in vivo imaging of the mouse brain at depths approaching 1 mm using 1035 nm excitation. C-FOCUS enables a range of biological applications that have not been previously achievable with two-photon microscopy at this wavelength: (1) visualization of axons in the white matter of the intact brain, demonstrating improved imaging depth while maintaining high resolution; (2) imaging of cerebral blood vessels up to 910 $\mu$m deep over a $252 \times 252~\mu m^2$ FOV with subregion correction, resulting in more than a 20-fold increase in fluorescence signal in vivo; and (3) visualization of apical dendrites up to 520 $\mu$m deep in the brain through the 110 $\mu$m-thick intact skull of adult mice. To our knowledge, this is the first in vivo demonstration of intensity-based scattering correction for two-photon microscopy.

Across these demonstrations, the effectiveness of C-FOCUS varies, with fluorescence intensity improvements ranging from 2.3-fold (Fig.~\ref{fig3}f) to 61.3-fold (Extended Data Fig.~\hyperref[secA2]{2}f). The primary factor affecting correction performance is the spatial distribution of the fluorescent or SHG signal. C-FOCUS is more effective when signals are sparsely distributed in 3D, such as fluorescent beads (Fig.~\ref{fig2}) and blood vessels (Fig.~\ref{fig4}), compared to densely labeled structures (Fig.~\ref{fig3}, \ref{fig5}) or SHG signals (Extended Data Fig.~\hyperref[secA3]{3}). Other factors influencing correction effectiveness include the scattering properties of the tissue, the SNR of measurements under random pattern modulation, and the hyperparameters of the compressive sensing algorithm. When scattering is weak and does not significantly degrade fluorescence (Fig.~\ref{fig3}f), the limited dynamic range of the PMT constrains the achievable improvement. Conversely, when scattering is too strong, it not only reduces the overall SNR of fluorescence under random pattern illumination but also suppresses the variation across different patterns, making the fluorescence responses nearly indistinguishable. This lack of contrast hampers the reconstruction of high-frequency components in the correction mask, thereby limiting performance. The ultimate depth limit of C-FOCUS is determined by the transport mean free path.

C-FOCUS is designed to enhance laser power at the focal point rather than to improve resolution, which represents a fundamental difference from adaptive optics. While C-FOCUS achieves sufficient resolution to visualize dendrites and axons at depths of 900 $\mu$m in the brain, it does not aim for synaptic resolution. Due to the nature of tissue scattering, which effectively acts as a low-pass filter, achieving synaptic resolution at these depths remains a significant challenge. One potential solution is to integrate C-FOCUS with adaptive optics for complex field modulation \cite{Sohmen2024-qm,Xue2022-yr} or combine it with post-processing techniques \cite{Belthangady2019-vc,Xue2024-pa, Zheng2021-sf,Xue2018-ay, Wei2023-km, Xue2022-rf, Li2021-jh,Qu2024-xt, Qiao2024-dv, Cao2025-ps} to further enhance resolution and SNR.

Since the total correction time per $z$-plane typically ranges from 44 to 90 seconds, and the actual duration of continuous acquisition is only about 1.5 seconds per correction mask, no motion artifacts were observed during the process. Although dynamic scattering caused by blood flow changes rapidly \cite{Qureshi2017-on}, C-FOCUS corrects for the average scattering profile rather than instantaneous fluctuations. While regions with dense vasculature exhibit stronger scattering than those with sparse vasculature, the dominant factor affecting overall scattering is steady-state scattering from highly scattering structures such as calcium hydroxyapatite in the skull and myelin in the white matter. These components reduce the EAL far more significantly than dynamic blood flow (Extended Data Fig.~\hyperref[secA1]{1}). In summary, C-FOCUS is robust to physiological motion and dynamic scattering in in vivo imaging.

C-FOCUS is compatible with a range of strategies to further extend imaging depth and broaden its applicability. For example, it can be combined with longer excitation wavelengths (e.g., 1100-1617 nm) and red-shifted fluorophores, which have been shown to improve imaging depth in two-photon microscopy \cite{Kondo2017-lh, Liu2019-ae, Cheng2019-zy}. It may also enhance three-photon microscopy \cite{Wang:20, Qin2022-ns,Ouzounov2017-fk, Streich2021-wf}, where higher-order nonlinearity could yield even greater improvements. Beyond imaging, C-FOCUS offers potential for two-photon optogenetics \cite{Packer2012-mz, Prakash2012-bv,Zhang2018-ji, Adesnik2021-tm} by enabling spatially selective modulation of excitation power through positive or negative masks \cite{ruan2021optical,Zhang2017-ow}, leveraging the optical memory effect. In summary, C-FOCUS demonstrates the feasibility and effectiveness of intensity-based scattering correction for two-photon microscopy, enabling deep, large-scale, dendrite-resolving imaging in vivo, and holds promise as a versatile tool for neurobiology, immunology, and cancer biology.

\section*{Methods}\label{Methods}

\textbf{Animals} 

All experimental and surgical protocols were approved by the University of California, Davis, Institutional Animal Care and Use Committee. Both males and females were used in the experiments, with ages ranging from 5-11 weeks old. Thy1-YFP-H transgenic mice were used for imaging neurons and C57BL/6J wild-type mice were used for imaging blood vessels (Jackson Laboratories). Details on animal preparations are described below.  

\noindent\textbf{Optical setup}

C-FOCUS uses a femtosecond pulsed laser at 1035 nm wavelength and 1 MHz repetition rate (Monaco 1035-40-40 LX, Coherent). The maximum power used for C-FOCUS is 2.4 W (The maximum total power of the laser is 40 W but 37.6 W is used to pump an optical parametric amplifier, which is not used for C-FOCUS). A polarizing beam splitter cube (PBS123, Thorlabs) and a half-wave-plate (WPHSM05-1310) mounted on a motorized rotation mount (K10CR2, Thorlabs) are used to adjust the input power to the following optics in the system. Two 1-axis galvo mirrors (QS7X-AG, QS7Y-AG, Thorlabs) are placed on the Fourier plane to scan the laser beam in 2D, relayed by a 1:1 4-$f$ system (AC508-250-C-ML, ACT508-250-C-ML, Thorlabs). The laser beam is then expended with a 4-$f$ system (LA1401-B, AC508-100-C-ML, Thorlabs). Next, the laser beam is pre-dispersed by a ruled grating (GR13-0310, 300/mm, 1000 nm blaze, Thorlabs) on the Fourier plane to compensate for the dispersion induced by the DMD (DLP650LNIR, 1280$\times$800 pixels, maximum pattern rate 12.5 kHz, VIALUX), relayed by a 4-$f$ system (2 AC508-200-C-ML lenses, Thorlabs). The beam from the DMD is relayed to the back aperture of the objective lens by two 4-$f$ systems (AC508-150-C-ML, ACT508-300-C-ML; 2 ACT508-200-B-ML lenses, Thorlabs). A dichroic mirror (DMSP680B, Thorlabs) is used to reflect the beam to the back-aperture of the objective lens. An XLUMPlanFL N (NA 1.0, $\times$20, Olympus) is used for all figures except Fig.~\ref{fig1}d, which uses an XLPLN25XSVMP2 (NA 1.0, $\times$25, Olympus). The objective lens is mounted on a motorized $z$-stage (V-308, Physik Instrumente) for axial scanning. In the emission path, a shortpass filter (ET750sp-2p8, CHROMA) is used to block the reflected excitation light. A bandpass filter (AT535/40m, CHROMA, Fig.~\ref{fig1} and Fig.~\ref{fig3}-\ref{fig5}; AT635/60m, CHROMA, Fig.~\ref{fig2}) is used to transmit the fluorescence emission. The fluorescence is detected by a PMT (H15460, Hamamatsu). The  system is controlled by a computer (OptiPlex 5000 Tower, Dell) using MATLAB and a data acquisition card (PCIe-6363, X series DAQ, National Instruments) for signal input/output.

\noindent\textbf{Typical parameters for scattering correction}

\textbf{\textit{Data acquisition.}} The first step in C-FOCUS is to record fluorescence intensity while projecting random binary patterns. Each pattern consists of 100 $\times$ 100 superpixels, with each superpixel made up of 8 $\times$ 8 pixels. The random binary patterns are generated from a uniform discrete distribution, with 40\% of the superpixels turned on. For each correction mask (100 $\times$ 100 superpixels), 3000 random patterns are used in the experiments shown in Fig.~\ref{fig1}, Fig.~\ref{fig3}-\ref{fig5}, and these patterns are preloaded onto the DMD. The number of patterns used in Fig.~\ref{fig2} is specified in the main text. The projection time per random pattern is 0.5 ms (2 kHz) for Fig.~\ref{fig1}, Fig.~\ref{fig3}, and Fig.~\ref{fig5}; 1 ms (1 kHz) for Fig.~\ref{fig2} and Fig.~\ref{fig4}b-c; and 2 ms (0.5 kHz) for Fig.~\ref{fig4}d. Of each projection cycle, 0.1 ms is allocated for pattern switching while the laser is off, based on the DMD's maximum switching speed of 12.5 kHz. Since C-FOCUS is a single-shot method with no iterative steps, the total data acquisition time per correction mask (with 3000 patterns) ranges from 1.5 to 6 seconds, depending on the projection rate. Additional time for data transfer and system overhead ranges from 1.3 to 3.3 seconds per mask. Overall, the acquisition time is primarily determined by the projection time per random pattern. 

For point-scanning images without correction or with global correction (i.e., a single correction mask applied to the entire FOV, Fig.~\ref{fig2}), either a blank screen or the global correction mask is projected on the DMD while the galvo mirrors scan across the FOV at a 4 kHz rate. Each image consists of 288$\times$288 pixels per $z$-plane, corresponding to 0.44 $\mu$m/pixel, with a 10 $\mu$m step size between adjacent $z$-planes. For point-scanning images with subregion corrections, DMD projection is synchronized with the galvo mirrors to display the appropriate correction masks at their corresponding subregions. Since the DMD uses a fixed projection time for all frames within a preloaded sequence, the correction masks for the 8$\times$8 patches are arranged into a sequence of 8$\times$288 frames, with each subregion's mask duplicated accordingly (see details in the Supplementary information). The projection time is 9 ms per frame, with an additional 0.18 ms buffer for switching frames (during which the laser is off). The total exposure time is kept constant between corrected and uncorrected acquisitions to ensure a fair comparison.

\textbf{\textit{Content-aware subregions.}}
C-FOCUS initially divides the entire FOV ($126 \times 126~\mu m^2$ for neuronal imaging or $252 \times 252~\mu m^2$ for vascular imaging) into an $8 \times 8$ grid of patches. Within each patch, the local peak intensity is identified and selected as the target point. If the target within a patch is too dim or located too close to a neighboring target (with a minimum separation threshold of 20 $\mu$m for neurons and 30-50 $\mu$m for blood vessels), the patch is merged with adjacent patches to form a content-aware subregion. The highest peak intensity among the merged patches is then designated as the target for that subregion.

\textbf{\textit{Computing correction masks.}} The correction masks are calculated using a classic gradient descent algorithm, Fast Iterative Shrinkage-Thresholding Algorithm (FISTA) \cite{Beck2009-hn}, running on a CPU-based computer (OptiPlex 5000 Tower, Dell). The typical optimization hyperparameters are: a step size of $2\times10^{-8}$, TV regularizer weight $\alpha~=~0.008$, and a maximum of 1,000 iterations. For each correction mask, the input to the algorithm consists of fluorescence measurements under random pattern modulation (3000$\times$1 vector) and the corresponding binary random patterns ($3000\times100\times100$). The output is a grayscale correction mask of size $100\times100$ superpixels. The grayscale correction mask is then binarized using a DMD output-to-input power ratio of 30\% in all figures except Fig.~\ref{fig2}, for which the ratio is specified in the main text. The initial guess is generated using the 2P-FOCUS algorithm \cite{Zepeda2025-cm}, applied to the same 3000 undersampled measurements. For Fig.\ref{fig1}, \ref{fig3}, and \ref{fig5}, the computation time per correction mask ranges from 2.5 to 3.1 seconds. For Fig.\ref{fig4}, the convergence is slower due to lower SNR in the raw measurements and a reduced $\alpha$ value, resulting in computation times of 10-18.7 seconds per mask. The computation speed could be further improved using a GPU to parallelize processing across all subregions.

\noindent\textbf{Digital image processing}

Digital image processing was performed using MATLAB, ImageJ, and Imaris Viewer. Raw images were denoised using a notch filter and a median filter, and background subtraction was carried out using unsharp masking. For images acquired with subregion correction, patch boundaries were refined by locally correcting uneven intensity near the margins, typically within a 1-3 pixel border. Volumetric renderings were generated in Imaris Viewer, using a gamma value of 0.5-1.0 and a saturation range of 0.7-0.8 for visualization purposes. ImageJ was used to display 2D images with a ``Green Hot” colormap and to adjust image contrast.

\noindent\textbf{Beads sample preparation}

A suspension of red fluorescent beads (R700, Thermo Fisher Scientific, MA) was mixed with PDMS (Sylgard 184, Dow Inc., MI) and covered with a coverslip. Air bubbles in the mixture were removed using a vacuum desiccator. The sample was then cured by heating at 100 $^\circ$C for 35 minutes. After curing, a dissected mouse skull (about 3 mm in diameter) was affixed to the top of the coverslip using superglue (Gorilla).

\noindent\textbf{Mouse preparation for neuronal imaging through a cranial window}

Survival surgery for craniotomy was performed to enable neuronal imaging in the primary somatosensory cortex (Fig.~\ref{fig1}d) or visual cortex (Fig.~\ref{fig1}b, Fig.~\ref{fig3}) of the mouse brain. During surgery, the mouse was anesthetized using 2-3\% isoflurane delivered via a vaporizer. The head was stabilized using ear bars and a tooth bar on a stereotactic frame (51730, Stoelting). Toe pinch reflexes were monitored throughout the procedure to assess anesthetic depth. The mouse was maintained on a heating pad (53800, Stoelting) to sustain a body temperature of 37 $^\circ$C, monitored via a rectal probe. Ophthalmic ointment was applied to protect the eyes. Fur over the surgical area was removed, and the scalp was sterilized with alternating wipes of chlorhexidine and ethanol. The scalp was then excised to expose the skull. The periosteum was removed with a scalpel blade. A 3-mm-diameter craniotomy site was marked using a biopsy punch, and the skull at the site was gradually thinned and removed with a micromotor drill (51449, Stoelting), ensuring the dura remained intact. A cranial window was prepared prior to surgery by bonding two coverslips (3 mm and 5 mm diameter) with ultraviolet-cured optical adhesive (Norland NOA 65). The window was placed over the craniotomy and secured with dental cement (Parkell C\&B Metabond), which was also applied to seal all wound edges. Once the cement hardened, a stainless steel headplate was affixed using the same dental cement. Imaging was conducted at least one week after surgery. During imaging sessions, the mouse was anesthetized with 1.5-2\% isoflurane and head-fixed. 

\noindent\textbf{Mouse preparation for neuronal imaging through an intact skull}

The steps up to skull exposure followed the same procedure as described above for the craniotomy. After exposing the skull and removing the periosteum, a 5-mm coverslip was directly placed over the selected site (visual cortex in Fig.~\ref{fig5}) without thinning or removing the skull. The edge of the coverslip was glued to the skull using dental cement (Parkell C\&B Metabond). The remaining steps and the imaging session were performed as described in the previous section.

\noindent\textbf{Mouse preparation for blood vessel imaging through a cranial window}

Anesthesia was induced in wild-type mice using 2.0\% isoflurane in oxygen and maintained at 1.5–2.0\% during surgical preparation. Anesthetic depth was monitored via toe pinch and whisker movement. Body temperature was maintained at 37 $^\circ$C using a water-perfused thermal pad (Gaymar T/Pump). A custom-made metal plate was affixed to the skull, and a cranial window (1.5–2.5 mm in diameter) was created over the visual cortex for acute imaging. FITC-dextran (2.5\%, 1 g/40 mL in saline) was administered via retro-orbital injection. To minimize vibration during imaging, 1.1\% agarose in artificial cerebrospinal fluid at 37 $^\circ$C was applied over the cranial window. Acute imaging was performed immediately after surgery. During imaging, the mouse was kept under anesthesia with 1.5–2.0\% isoflurane and head-fixed.

\backmatter

\bmhead{Author contributions}
Y. X. conceived and led the project, and designed and built the microscopy system. Y. X., R. H., and Y. L. conducted the experiments and processed the data. B. U. prepared the mouse used in Fig.~\ref{fig4} under the guidance of J. W.. Y. X. wrote the paper with inputs from all authors.

\bmhead{Acknowledgments}
We thank Dr. Hillel Adesnik and Dr. Kevin Sit at the University of California, Berkeley, for their guidance on mouse craniotomy survival surgery and for providing the PMT used in the microscopy setup. We also thank Dr. Weijian Yang and Ben Mattison at UC Davis for sharing the design of the mouse headplate. Research reported in this publication was supported by the National Institute of General Medical Sciences of the National Institutes of Health 1R35GM155193-01. The content is solely the responsibility of the authors and does not necessarily represent the official views of the National Institutes of Health. This work was also supported by Dr. Yi Xue’s startup funds from the Department of Biomedical Engineering at the University of California, Davis.

\bmhead{Data Availability}
The main data supporting the findings of this study are available within the paper and its Supplementary information files. The source data files for all data presented in this paper can be found at: \url{https://xue-lab-cobi.github.io/c_focus/}.

\bmhead{Code Availability}
The code for correction mask calculation and digital image processing are available at: \url{https://xue-lab-cobi.github.io/c_focus/}.

\section*{Declarations}
The authors declare no competing interests.

\bibliography{sn-bibliography}

\begin{appendices}
\section{Extended Data Fig.1}\label{secA1}
\begin{figure}[h!]%
\centering
\includegraphics[width=1\textwidth]{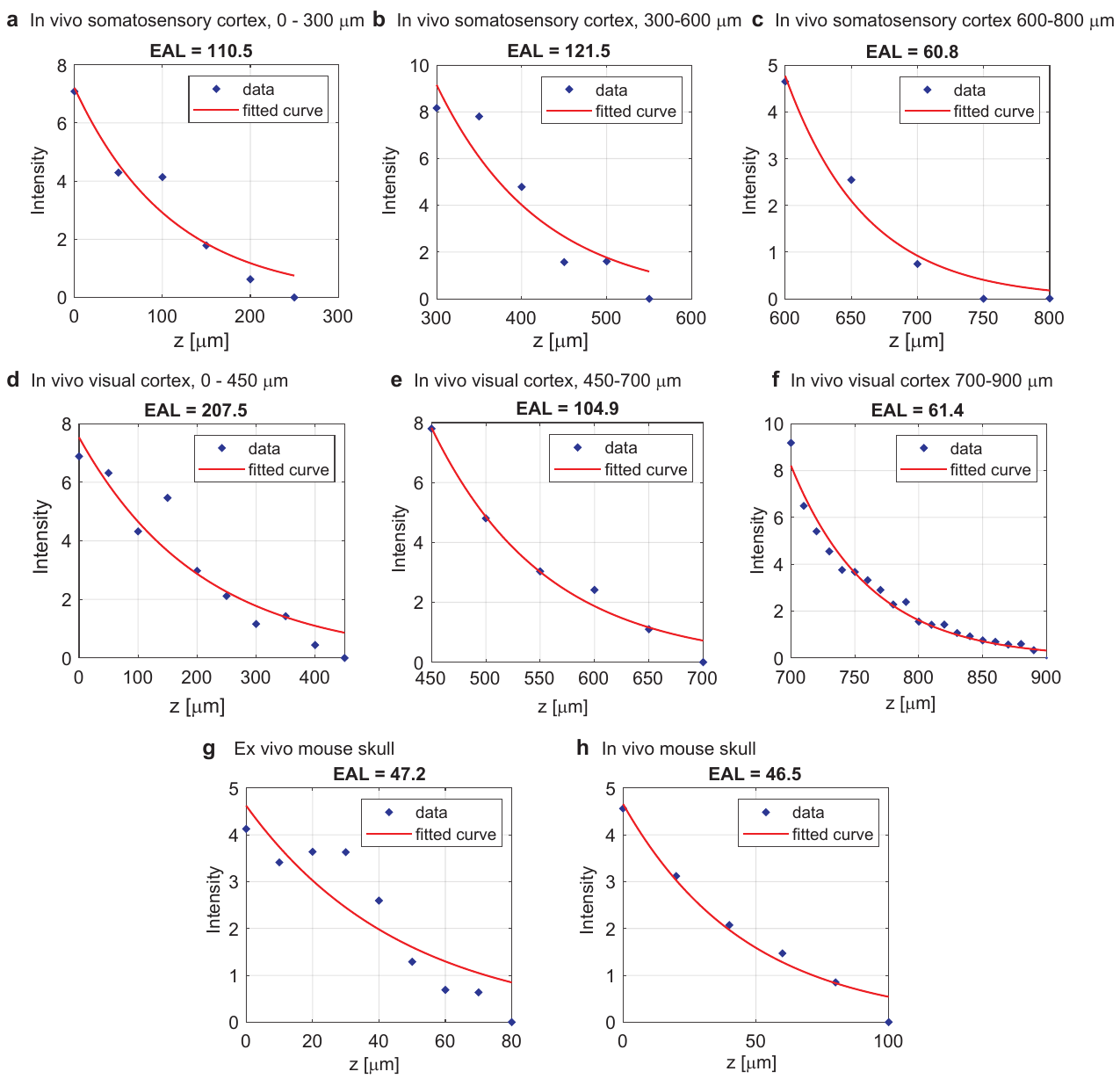}
\caption{\textbf{Experimentally measured effective attenuation length (EAL) in various biological samples.} The EAL is calculated by fitting an exponential decay curve $I_z = I_0 e^{-z/EAL}$, where $I_z$ and $I_0$ are the estimated laser intensities (a.u.) at depth $z$ and at the initial plane, respectively, calculated by $I = \sqrt{I_{fluo}}$. \textbf{a-c.} In vivo EALs in the primary somatosensory cortex of a Thy1-YFP-H mouse are: \textbf{a.} 110.5 $\mu$m at 0-300 $\mu$m depth, \textbf{b.} 121.5 $\mu$m at 300-600 $\mu$m depth, and \textbf{c.} 60.8 $\mu$m at 600-800 $\mu$m depth. Severe scattering near the brain surface is caused by pial arteries, while increased scattering in deeper layers is attributed to myelinated axons. 
\textbf{d-f.} In vivo EALs in the visual cortex of a Thy1-YFP-H mouse are \textbf{d.} 207.5 $\mu$m at 0-450 $\mu$m depth, \textbf{e.} 104.9 $\mu$m at 450-700 $\mu$m depth, and \textbf{f.} 61.4 $\mu$m at 700-900 $\mu$m depth. Compared to the primary somatosensory cortex, the visual cortex is thinner and has less dense vasculature. Scattering is primarily caused by myelinated axons in the deep cortical layers and underlying white matter. 
\textbf{g.} EAL of the ex vivo mouse skull (Fig. 2) is 47.2 $\mu$m at 0-80 $\mu$m depth. \textbf{h.} EAL of the in vivo mouse skull (Fig. 5) is 46.45 $\mu$m at 0-100 $\mu$m, similar to that of the ex vivo skull. }\label{figA1}
\end{figure}

\section{Extended Data Fig.2}\label{secA2}
\begin{figure}[h!]%
\centering
\includegraphics[width=1\textwidth]{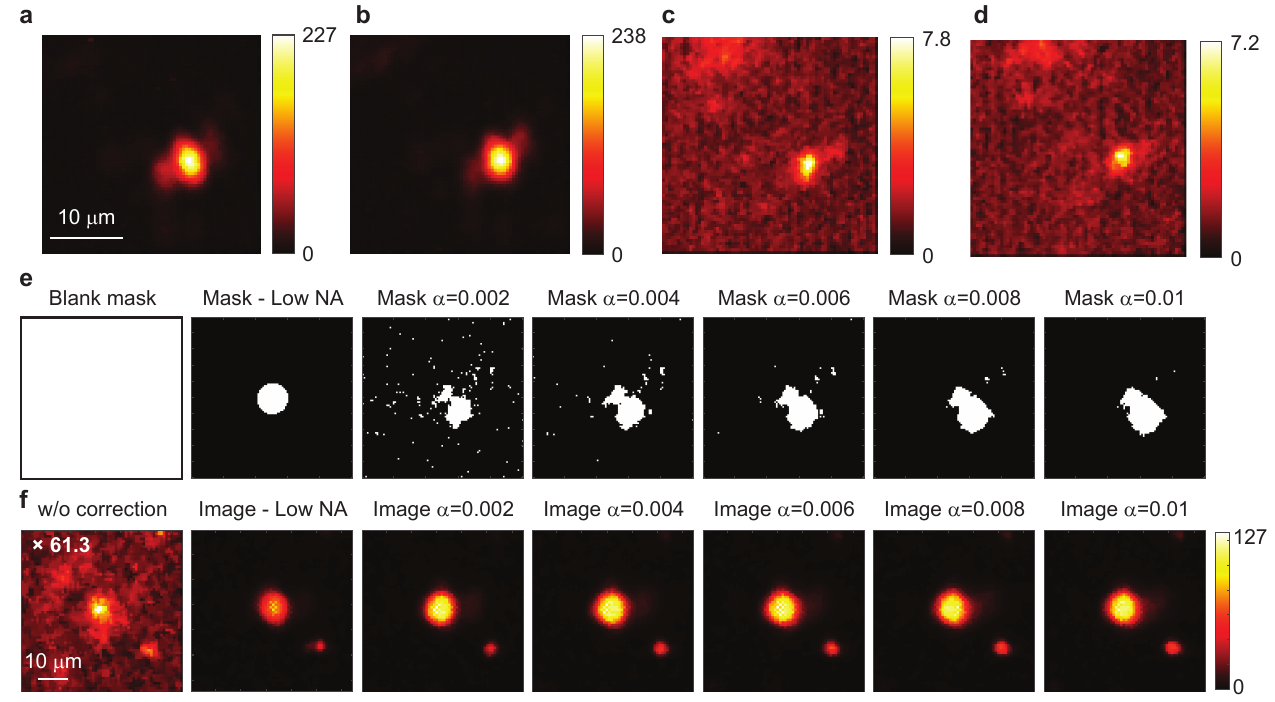}
\caption{\textbf{Imaging fluorescent beads through ex vivo mouse skull. a-b.} Fluorescence intensity improvement is independent of the choice of random pattern seeds. \textbf{a.} Fluorescence intensity with a correction mask generated using 2,000 random patterns. \textbf{b.} Fluorescence intensity with a correction mask generated using a different set of 2,000 random patterns. The random seed was reshuffled to generate the two sets, with all other parameters held constant. \textbf{c-d.} Fluorescence intensity without correction measured \textbf{c.} before and \textbf{d.} after all imaging tasks, indicating that photobleaching was negligible and intensity comparisons are valid regardless of imaging sequence. \textbf{e-f.} Correction effectiveness varies with the hyperparameters of the compressive sensing algorithm. \textbf{e.} Blank mask, low-NA mask, and correction masks computed using different weights of the TV regularizer ($\alpha$) in Eq.\ref{eq1}. \textbf{f.} Images of fluorescent beads through bone acquired using the corresponding masks in \textbf{e}. The differences between masks with different $\alpha$ values are minor (less than 4\%), with the maximum fluorescence intensity achieved at $\alpha = 0.006$, resulting in a 61.3-fold improvement compared to the case without scattering correction.}\label{figA2}
\end{figure}

\section{Extended Data Fig.3}\label{secA3}
\begin{figure}[h!]%
\centering
\includegraphics[width=1\textwidth]{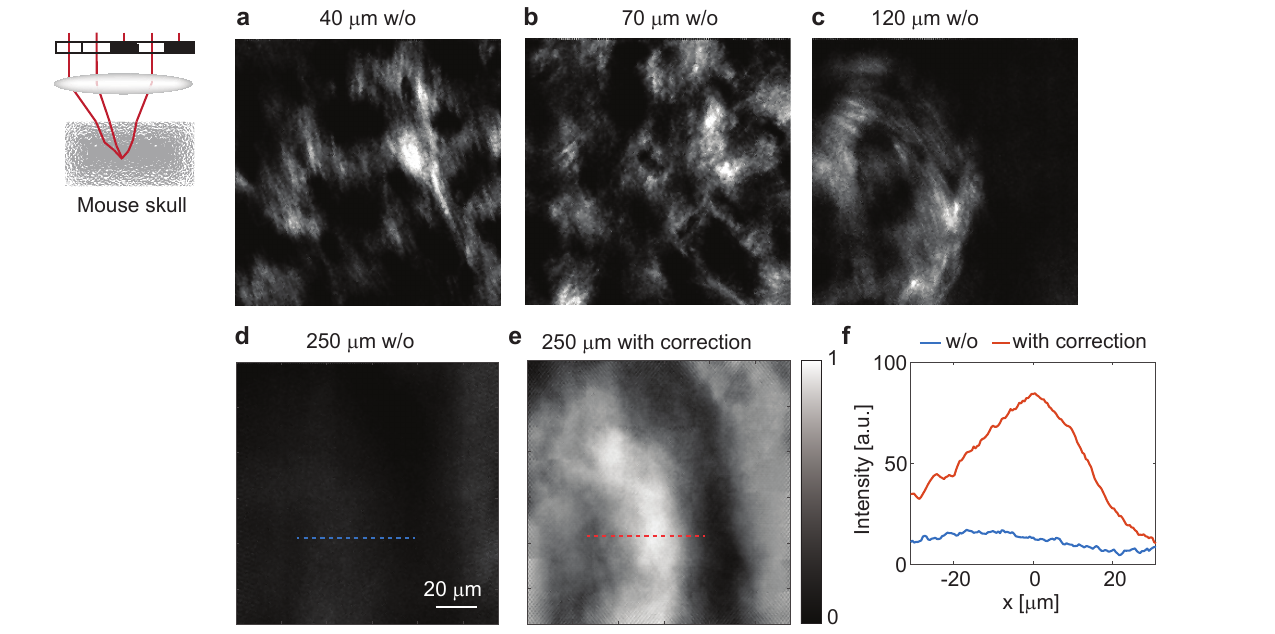}
\caption{\textbf{Imaging SHG signals in the ex vivo mouse skull.} The skull was dissected from an aged mouse, with an EAL of 47.2 $\mu$m (Extended Data Fig.~\hyperref[secA1]{1}). The maximum laser power on the sample surface is 22 mW. \textbf{a-d.} SHG images without correction at depths of \textbf{a.} 40 $\mu$m, \textbf{b.} 70 $\mu$m, \textbf{c.} 120 $\mu$m, and \textbf{d.} 250 $\mu$m. \textbf{e.} SHG image of the same region shown in \textbf{d} with correction. \textbf{f.} Comparison of the SHG signal along the dashed line in \textbf{d} and \textbf{e}, showing a significant improvement in the SHG signal with correction. }\label{figA3}
\end{figure}

\end{appendices}


\end{document}